\documentclass[aps,amsmath,notitlepage,twocolumn,amssymb,prx,longbibliography]{revtex4-1}

\usepackage{times}
\usepackage{graphicx}
\usepackage{float}
\usepackage{latexsym,amsmath,amssymb,bm,euscript}
\usepackage{color}
\usepackage{subfigure}
\usepackage{epstopdf}
\usepackage[colorlinks=true,linkcolor=blue,citecolor=blue]{hyperref}
\usepackage{soul}
\usepackage[normalem]{ulem}
\usepackage{mathrsfs}
\usepackage{amsmath}
\usepackage{lettrine}
\usepackage{bbding}
\usepackage{xspace}
\usepackage{textcomp}
\usepackage{textcase}
\usepackage{setspace}
\usepackage{multirow}
\usepackage[version=4]{mhchem}

\def\para{\ensuremath{/\kern -0.8em /}\xspace}

\def\beqn{\begin{eqnarray}}
\def\eeqn{\end{eqnarray}}
\def\beq{\begin{equation}}
\def\eeq{\end{equation}}

\newcommand{\Beq}{\begin{eqnarray*} }
\newcommand{\Eeq}{\end{eqnarray*} }
\newcommand{\Bmat}{\left(\begin{matrix}}
\newcommand{\Emat}{\end{matrix}\right)}

\begin{document}

\title{Emergent Dispersive Multipolar Excitations in \ce{NaErSe2}}

\author{Zheng Zhang$^{1}$}
\author{Mingfang Shu$^{2,5}$}
\author{Mingtai Xie$^{1,3}$}
\author{Weizhen Zhuo$^{1,3}$}
\author{Yanzhen Cai$^{1,3}$}
\author{Christian Balz$^{4}$}
\author{Jianting Ji$^{1}$}
\author{Feng Jin$^{1}$}
\author{Jie Ma$^{2}$}
\email{jma3@sjtu.edu.cn}
\author{Qingming Zhang$^{1,3}$}
\email{qmzhang@iphy.ac.cn}

\affiliation{$^{1}$Beijing National Laboratory for Condensed Matter Physics, Institute of Physics, Chinese Academy of Sciences, Beijing 100190, China}
\affiliation{$^{2}$Department of Physics and Astronomy, Shanghai Jiao Tong University, Shanghai 200240, China}
\affiliation{$^{3}$School of Physical Science and Technology, Lanzhou University, Lanzhou 730000, China}
\affiliation{$^{4}$ISIS Facility, STFC Rutherford-Appleton Laboratory, Didcot, OX11 0QX, United Kingdom}
\affiliation{$^{5}$College of Sciences, China Jiliang University, Hangzhou 310018, China}

\begin{abstract}
In most condensed-matter systems, local and collective excitations remain decoupled due to their distinct energy scales. 
Here, we identify the coupled local–collective excitations in the triangular antiferromagnet \ce{NaErSe2}, by combining neutron spectroscopy with a total angular momentum modeling. 
The low-lying crystalline electric field (CEF) doublets include a dipolar $\Gamma_4$ ground state forming a stripe-$x$ order and a $\Gamma_{5,6}$ excited state with dipole–octupole character. 
High-resolution spectra reveal emergent symmetry-selected
dispersions where magnon branches from the ground state are replicated on higher $\Gamma_4$ levels but couple with the $\Gamma_{5,6}$ levels to form a distinct multipolar band. 
An applied magnetic field reconstructs the CEF wavefunctions and polarizes the system into a multipolar ferromagnet, further reshaping the spectra. 
The study demonstrates the emergent coupling of local and collective excitations driven by strong spin-orbit coupling and establishes \ce{NaErSe2} as a platform for field-tunable multipolar excitations in frustrated magnets.
\end{abstract}
\date{\today}
\maketitle

\textit{Introduction}-----
Rare-earth magnets, characterized by their strongly localized 4$f$ electrons and suppressed charge fluctuations, offer a unique platform for exploring emergent quantum phenomena in strongly correlated systems.
The strong spin–orbit coupling (SOC) inherent to rare-earth ions entangles spin and orbital degrees of freedom, giving rise to well-defined total angular momentum states labeled by $J$~\cite{Stevens_1952}.
Within a crystalline electric field (CEF), the SOC-locked $J$ manifold is further split into multiplets that carry not only dipolar moments but also a hierarchy of higher-rank multipolar components\cite{PhysRevB.26.5451,PhysRevB.27.7386,RevModPhys.82.53,PhysRevLett.112.167203}. 
These ingredients underpin a diverse range of magnetic ground states and excitation spectra, including dipolar quantum spin liquids (QSL)~\cite{paddison_continuous_2017,PhysRevLett.115.167203,PhysRevLett.117.097201,bordelon_field-tunable_2019,PhysRevB.100.241116,zhuo_magnetism_2024,PhysRevX.11.021044,PhysRevB.106.085115}, spin ice~\cite{PhysRevLett.122.187201,doi:10.1126/science.1178868,doi:10.1126/science.1177582}, U(1) spin liquid~\cite{PhysRevLett.115.097202,PhysRevB.101.144419,sibille_quantum_2020,gao_experimental_2019,PhysRevResearch.2.013334}, unconventional multipolar orders~\cite{PhysRevB.94.201114,PhysRevB.95.041106,PhysRevB.98.045119}, transverse Ising model~\cite{PhysRevResearch.2.043013,RN109,RN116,RN122,PhysRevB.111.L180405,PhysRevB.108.054435,PhysRevB.109.075159}, and Kitaev-related excitations~\cite{PhysRevLett.133.096703,PhysRevResearch.7.023198}.

A central challenge in this context is to understand and control the coupling between local CEF excitations and collective spin dynamics --- a regime rarely accessed in conventional systems due to the disparity in energy scales.
Yet in certain rare-earth compounds, notably those based on Er$^{3+}$ with its large $J = 15/2$, the energy scales of CEF splittings and exchange interactions become comparable, enabling such coupling to occur naturally. 
This opens a pathway to discovering new types of dispersive modes beyond conventional magnons or local CEF excitations, especially when symmetry constraints and multipolar character come into play.

In this work, we report the discovery of symmetry-selected dispersive multipolar excitations in the triangular-lattice magnet \ce{NaErSe2}, a representative member of the rare-earth chalcogenide family~\cite{Liu_2018,cpl_41_11_117505}.
Using high-resolution inelastic neutron scattering (INS), we identify five low-lying CEF doublets below 6 meV, whose dispersions couple with magnon branches originating from a dipolar stripe-$x$ ground state. 
Of particular interest is the first excited $\Gamma_{5,6}$ dipole–octupole (DO) doublet, which --- despite being symmetry-distinct --- strongly couples to the ground-state magnons, producing a new dispersive multipolar band. 
This provides direct spectroscopic evidence for local–collective coupling, a phenomenon that lies beyond the conventional magnon or CEF picture.

To capture this physics, we go beyond effective spin-1/2 models and employ a total angular momentum $J=15/2$ manifold approach, in which both CEF and exchange terms are treated on equal footing. 
This unified framework quantitatively reproduces not only the zero-field excitation spectra, but also their evolution under applied magnetic fields. 
Importantly, the success of this modeling links the extracted CEF parameters to the effective exchange energy scale, explaining why such CEF–magnon coupling is particularly prominent in Er-based systems.

Finally, by applying magnetic fields along the $a$-axis, we demonstrate a field-induced crossover from a dipolar stripe phase to a ferromagnetic state. 
This transition is accompanied by a reconstruction of the wavefunctions that injects higher-rank multipoles into the ordered state, leading to a global reshaping of the excitation spectrum. 
Taken together, our study positions \ce{NaErSe2} as a prototype for a broader class of rare-earth magnets with weak CEF splittings, demonstrating a unified $J$-manifold framework that simultaneously captures dipolar stripe order and the emergence of dispersive multipolar excitations. 
By showing how symmetry, strong SOC, and external fields cooperate to tune both the ground state and the multipolar spectrum, we provide a roadmap for engineering novel magnetic phases and excitation modes across the rare-earth family—opening avenues for field-controlled functionality, from exotic quasiparticles to magnetocaloric applications.

\begin{figure}[t!]
	\includegraphics[angle=0,width=1\linewidth]{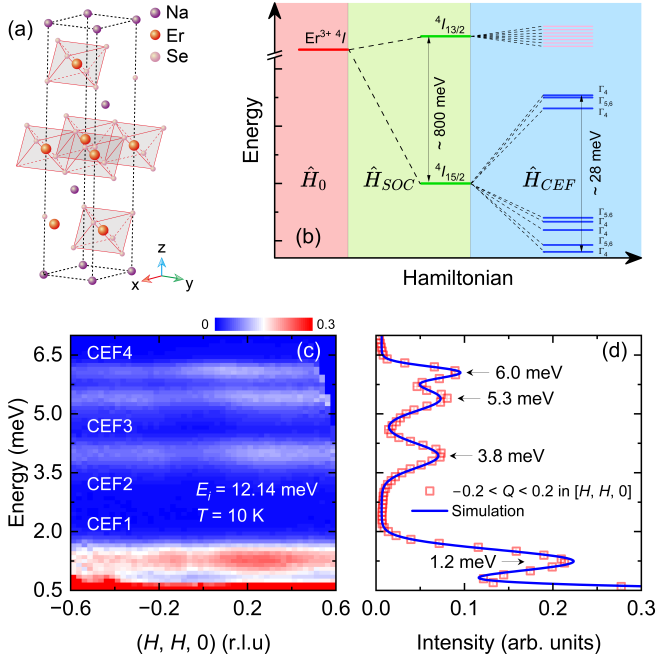}
	\renewcommand{\figurename}{\textbf{Fig. }}
	\caption{\textbf{Crystal structure and CEF excitations in \ce{NaErSe2}.}
		\textbf{(a)} Crystal structure of \ce{NaErSe2}. 
		\textbf{(b)} Energy level hierarchy of the Er$^{3+}$ $4f^{11}$ manifold under the central potential ($\hat{H}_0$), SOC ($\hat{H}_{\mathrm{SOC}}$), and the CEF ($\hat{H}_{\mathrm{CEF}}$).
		\textbf{(c)} INS spectrum at $T=10$~K, showing four distinct CEF excitations. 
		\textbf{(d)} Momentum-integrated energy cut of the data in \textbf{(c)}, compared with a CEF simulation (solid blue line).
		}
	\label{Fig1:CEF}
\end{figure}

\textit{CEF excitations of \ce{NaErSe2}}---In a central potential $\hat{H}_0$, the Er$^{3+}$ ion adopts a $4f^{11}$ configuration, with SOC splitting it into multiples. The ground state $^4I_{15/2}$ lies well below the $^4I_{13/2}$ manifold by $\sim$800 meV~\cite{Dieke63}, making it the only relevant manifold at low temperatures. The CEF, constrained by time-reversal symmetry, further splits this manifold into Kramers doublets,  Fig.~\ref{Fig1:CEF}\textbf{(b)}, reflecting the hierarchy of energy scales.

In \ce{NaErSe2}, each \ce{Er^{3+}} ion is octahedrally coordinated by six \ce{Se^{2-}} anions, forming a local CEF with $D_{3d}$ symmetry [Fig.~\ref{Fig1:CEF}\textbf{(a)}].  
Within the $J = 15/2$ manifold, the CEF Hamiltonian reads
\begin{equation}
	\hat{H}_{\mathrm{CEF}} = \sum_{i} B_{2}^{0} \hat{O}_{2}^{0} + B_{4}^{0} \hat{O}_{4}^{0} + B_{4}^{3} \hat{O}_{4}^{3} + B_{6}^{0} \hat{O}_{6}^{0} + B_{6}^{3} \hat{O}_{6}^{3} + B_{6}^{6} \hat{O}_{6}^{6}
	\label{Eq:CEFHamiltonian}
\end{equation}
where $B_{m}^{n}$ are the CEF parameters and $\hat{O}_{m}^{n}$ are Stevens operators constructed from the total angular momentum $\hat{J}$.

To probe the CEF excitations in \ce{NaErSe2}, we performed INS measurements at the LET spectrometer at the ISIS Facility~\cite{neutron,SM}, revealing four distinct modes at 1.2, 3.8, 5.3, and 6.0~meV [Fig.~\ref{Fig1:CEF}\textbf{(c)}], consistent with values for related compounds~\cite{PhysRevB.101.144432,PhysRevB.102.024424}. 
We quantified this scheme by analyzing $\vec{Q}$-integrated cuts of the data [Fig.~\ref{Fig1:CEF}\textbf{(d)}] using a standard CEF model~\cite{Scheie:in5044}. 
The robustness of this model is comprehensively validated by two independent probes: Raman spectroscopy, which confirms the low-energy modes and observes predicted higher-energy levels up to $\sim$28~meV, and bulk magnetization, which is accurately reproduced by including a mean-field correction for exchange interactions. 
The complete CEF scheme is detailed in the Supplemental Material (SM)~\cite{SM}.

A richer physical picture can be obtained from a symmetry analysis of the validated CEF wavefunctions. 
The analysis reveals that the CEF states split into two classes: the ground state, along with the second, third, fifth, and seventh levels, transform as the $\Gamma_4$ irreducible representation, while the first, fourth, and sixth levels transform as $\Gamma_{5,6}$.
This identifies the ground state as a purely dipolar doublet and the first excited state as a dipole-octupole (DO) doublet with significant multipolar character. This crucial distinction opens the door to exploring the interplay between dipolar order and multipolar excitations in \ce{NaErSe2}.

\begin{figure*}[t!]
	\includegraphics[angle=0,width=1\linewidth]{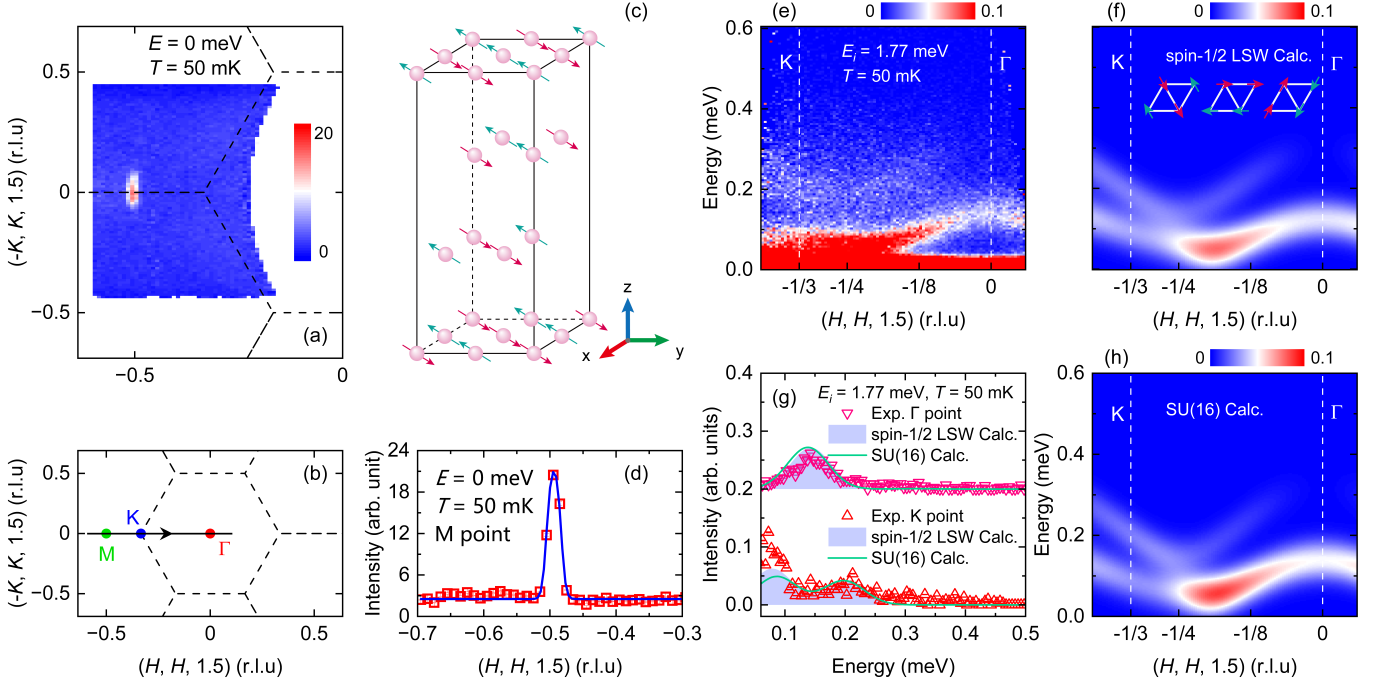}
	\renewcommand{\figurename}{\textbf{Fig. }}
	\caption{\textbf{Stripe-$x$ magnetic order and spin excitations in \ce{NaErSe2}.} 
		\textbf{(a)} Elastic neutron scattering intensity map at 50 mK, showing a magnetic Bragg peak at the M point. A corresponding intensity cut is shown in \textbf{(d)}. 
		\textbf{(b)} Reciprocal space diagram illustrating the high-symmetry points and measurement path. 
		\textbf{(c)} The determined stripe-$x$ magnetic structure. 
		\textbf{(e)} Measured low-energy spin-wave spectrum along the $\left(H, H, 1.5\right)$ direction. 
		This is compared with calculations using an effective spin-1/2 LSW model \textbf{(f)} and a $J = 15/2$ SU($N$) model \textbf{(h)} . 
		The inset in \textbf{(f)} shows the three degenerate stripe configurations. 
		\textbf{(g)} Constant-$\vec{Q}$ energy cuts at the M and $\Gamma$ points, comparing experimental data with both theoretical models.
	}
	\label{Fig2:GroundState}
\end{figure*}

\textit{Dipolar stripe-$x$ ground state}---While the CEF excitations are well characterized, a full understanding of \ce{NaErSe2} requires determination of its magnetic ground state and low-energy spin dynamics. Specific heat data exhibits a sharp anomaly near 0.2 K, signaling magnetic ordering (see SM~\cite{SM}).
Analysis of the CEF wave functions identifies a dipolar ground-state doublet, consistent with the magnetic Bragg peak observed at $\vec{Q} = (-0.5, -0.5, 1.5)$ at 50 mK [Fig.~\ref{Fig2:GroundState}\textbf{(a)} and \textbf{(d)}].
Accounting for lattice symmetry, this magnetic Bragg peak is consistent with stripe order characterized by $\vec{Q}$ = $\left(0.5, 0.5, 0.5\right)$.

Our INS on \ce{NaErSe2}~[Fig.~\ref{Fig2:GroundState}\textbf{(e)}] reveals well-defined spin-wave excitations, enabling a definitive determination of the magnetic ground state. 
To identify the specific ordered phase, we model the spin dynamics using an effective spin-1/2 anisotropic Hamiltonian, which provides a minimal yet physically meaningful description of the dipolar doublet manifold~\cite{PhysRevLett.115.167203,PhysRevB.94.035107}.
On a triangular lattice, this Hamiltonian is known to stabilize two competing stripe-ordered states~\cite{PhysRevB.94.035107,PhysRevX.9.021017,PhysRevB.103.205122}: 
an in-plane stripe-$x$ configuration and a stripe-$yz$ phase featuring a sizable out-of-plane moment component~\cite{PhysRevX.9.021017,PhysRevB.95.165110,PhysRevB.94.035107,PhysRevB.95.165110}.

By applying linear spin wave (LSW) theory~\cite{dahlbom2025sunnyjljuliapackagespin}, we find that a stripe-$x$  magnetic ground state provides an excellent description of the INS data for \ce{NaErSe2}.
As shown in Fig.~\ref{Fig2:GroundState}, the computed spectrum quantitatively reproduces the overall dispersion and the energy cuts at high-symmetry K and $\Gamma$ points.
The model further predicts a small gap of $\sim$ 0.04 meV along the $\left[H, H, 1.5\right]$ direction, which is slightly below but comparable to the instrumental resolution ($\Delta E \sim 0.05$ meV).
This excellent agreement is achieved with a single set of exchange parameters:  $J_{zz}$ $=$ $0.30$ K, $J_{\pm}$ $=$ $-0.10$ K, $J_{\pm\pm}$ $=$ $-0.63$ K, and $J_{z \pm}$ $=$ $0.29$ K (with an approximate error of 3.9\%).
The strong intrinsic anisotropy revealed by these parameters (e.g., $|J_{\pm\pm}/J_{\pm}| \approx 6.3$) explains why this nearest-neighbor (NN) only model is sufficient, in contrast to weakly anisotropic Yb-based materials, where the 120° magnetic order is readily destabilized, and a small next-nearest-neighbor ($J_{2}$) interaction of only $J_{2} \sim 0.05J_{1}$ is sufficient to promote a quantum spin liquid phase~\cite{bordelon_field-tunable_2019,PhysRevB.92.041105,steinhardt_phase_2021}.
Furthermore, this stripe-$x$ ground state is unambiguously confirmed by our independent density matrix renormalization group calculations~\cite{SM, ITensor, ITensor-r0.3}.
 
The small but finite gap, a consequence of the $C_{3}$ symmetry breaking, is a hallmark of an Order-by-Disorder (ObD) mechanism where strong anisotropic exchange resolves the inherent geometric frustration to select the stripe-ordered ground state~\cite{PhysRevX.9.021017}.
This scenario is analogous to the celebrated case of quantum ObD in the pyrochlore \ce{Er2Ti2O7}~\cite{PhysRevLett.112.057201,PhysRevLett.109.167201},  where quantum fluctuations lift an accidental U(1) degeneracy of the classical ground states at the mean-field level.
The general principle—that strong anisotropic interactions stabilize a specific stripe order from a frustrated manifold—is further exemplified in Ce-based magnets like \ce{CsCeSe2}~\cite{PhysRevLett.133.096703} and \ce{KCeSe2}~\cite{PhysRevResearch.7.023198}, where dominant Kitaev interactions similarly induce a stripe-$yz$ phase.

The effective spin-1/2 model accurately captures the ground-state spin dynamics of \ce{NaErSe2}. However, understanding the observed dispersive multipolar excitations, which involve higher-lying CEF states, requires a more comprehensive total angular momentum $J$ manifold approach. We now employ this framework, which naturally incorporates the coupling between local CEF and collective magnetic excitations, to analyze the full excitation spectrum.

\begin{figure}[t!]
	\includegraphics[angle=0,width=1\linewidth]{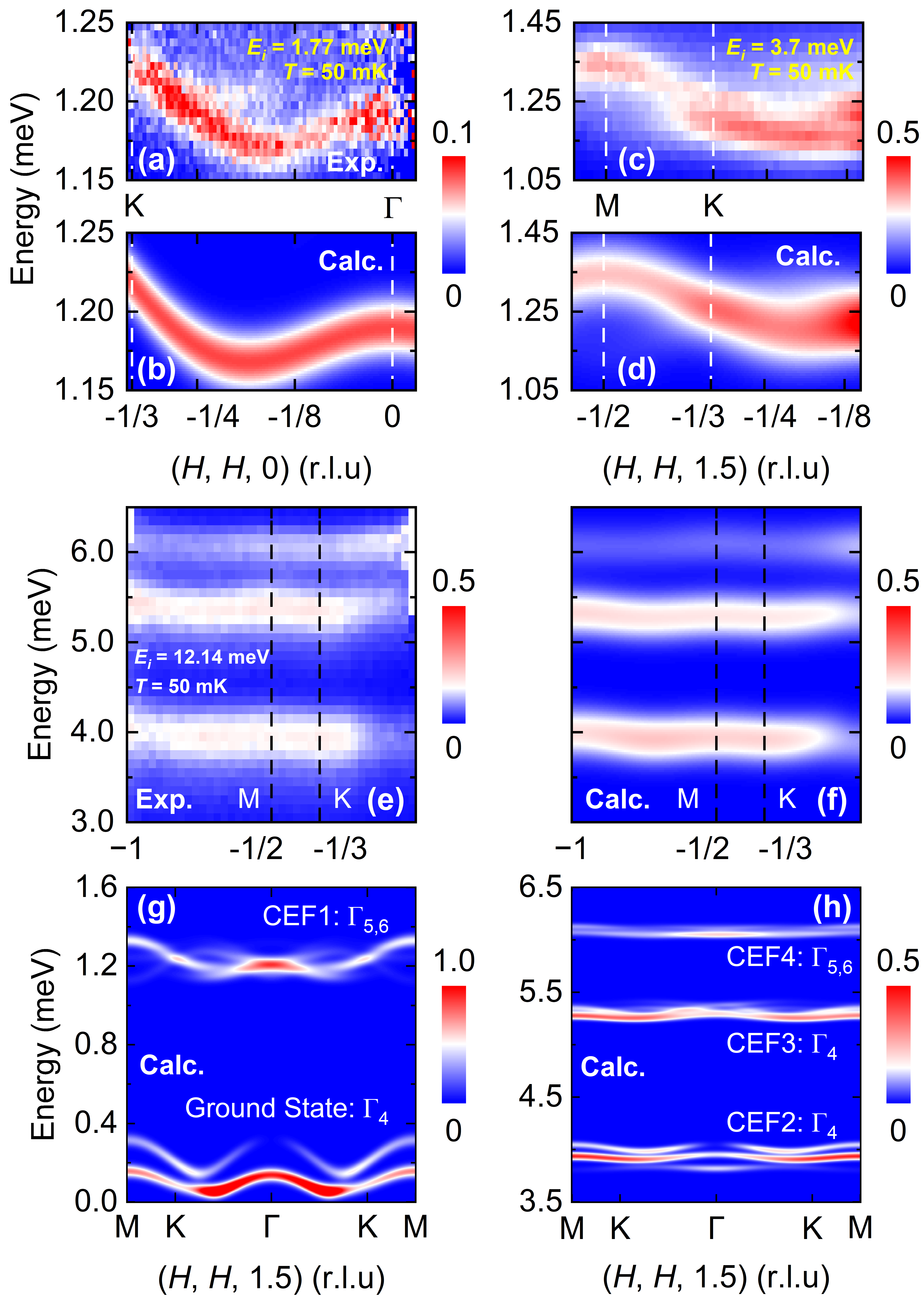}
	\renewcommand{\figurename}{\textbf{Fig. }}
	\caption{\textbf{Dispersive CEF excitations in \ce{NaErSe2}.} 
	\textbf{(a, c, e)} Experimental INS spectra measured at $T=50$ mK with incident energies $E_i=1.77$, $3.7$, and $12.14$ meV, respectively. 
	\textbf{(b, d, f)} Corresponding theoretical calculations using the $J=15/2$ SU($N$) model. 
	The momentum path is along $\left(H, H, 0\right)$ for \textbf{(a, b)} and along $\left(H, H, 1.5\right)$ for \textbf{(c-f)}.
	\textbf{(g, h)} High-resolution SU($N$) calculations showing the detailed dispersions of the ground state and the first four CEF excitations, labeled with their respective symmetry (e.g., $\Gamma_{4}$, $\Gamma_{5,6}$).
	}
	\label{Fig3:DisCEF}
\end{figure}

\textit{Dispersive multipolar excitations}---In conventional magnets, low-energy spin and CEF excitations are typically considered independent due to their well-separated energy scales. 
However, INS measurements on \ce{NaErSe2} with higher incident neutron energies reveal unexpected behavior.
The underlying stripe-$x$ order induces a pronounced dispersion in the low-lying CEF excitations [Fig.~\ref{Fig3:DisCEF}]. While this is most visually apparent for the first excited level in the INS spectra of Fig.~\ref{Fig3:DisCEF}\textbf{(a)} and \ref{Fig3:DisCEF}\textbf{(c)}, constant-energy cuts reveal a clear momentum dependence in the scattering intensity for the second and third levels as well (see SM~\cite{SM}).
This dispersion calls for a unified framework to describe the coupling of CEF and collective magnetic excitations.

To capture the coupling, we employ an effective Hamiltonian formulated within the total angular momentum $J$ manifold, $\hat{H}_{\mathrm{eff}} = \hat{H}_{\mathrm{CEF}} + \hat{H}_{\mathrm{exc}}$, where $\hat{H}_{\mathrm{CEF}}$ is given in Eq.~\ref{Eq:CEFHamiltonian}, and $\hat{H}_{\mathrm{exc}}$ describes the anisotropic magnetic interactions within the total angular momentum $J$ manifold.
Following the symmetry analysis of \ce{YbMgGaO4}~\cite{PhysRevLett.115.167203,PhysRevX.9.021017}, we write the $\hat{H}_{\mathrm{exc}}$ for \ce{NaErSe2} as~\cite{Liu_2024, SM}
\begin{align}
	& \hat{H}_{\mathrm{exc}} = \sum_{\langle i j\rangle} \left[ \vartheta_{z z} \hat{J}_i^z \hat{J}_j^z+ \vartheta_{ \pm}\left(\hat{J}_i^{+} \hat{J}_j^{-}+\hat{J}_i^{-} \hat{J}_j^{+}\right)\right. \notag \\
	& +\vartheta_{ \pm \pm}\left(\gamma_{i j} \hat{J}_i^{+} \hat{J}_j^{+}+\gamma_{i j}^* \hat{J}_i^{-} \hat{J}_j^{-}\right) \notag\\
	& \left.-\frac{i \vartheta_{z \pm}}{2}\left(\gamma_{i j} \hat{J}_i^{+} \hat{J}_j^z-\gamma_{i j}^* \hat{J}_i^{-} \hat{J}_j^z+\langle i \longleftrightarrow j\rangle\right) \right] 
	\label{Eq:HamiltonianEx}
\end{align}
Here, $\hat{J}_{i}^{\alpha}$ ($\alpha$ = $x, y, z$) denote the $J = 15/2$ total angular momentum operators at site $i$, and $\hat{J}_{i}^{\pm} = \hat{J}_{i}^{x} \pm i\hat{J}_{i}^{y}$ are the corresponding non-Hermitian ladder operators. The nearest neighbor anisotropic interactions are denoted by $\vartheta_{zz}$, $\vartheta_{\pm}$, $\vartheta_{\pm\pm}$, and $\vartheta_{z \pm}$. The phase factor $\gamma_{i j}$ is taken as 1, $e^{i\frac{2\pi}{3}}$, $e^{-i\frac{2\pi}{3}}$ along the three bonds $\vec{a}_{1}$, $\vec{a}_{2}$, and $\vec{a}_{3}$ (see SM~\cite{SM}).

Since the spin operators are consistently defined within the $J = 15/2$ manifold, the CEF and exchange Hamiltonians couple naturally in a unified framework. 
We implement the SU$(N)$ coherent states method~\cite{dahlbom2025sunnyjljuliapackagespin,PhysRevB.104.104409} to simulate both the ground state spin wave excitations and the dispersive CEF modes using the effective Hamiltonian $\hat{H}_{\mathrm{eff}}$.
Our calculations identified a set of exchange interactions --- $\vartheta_{zz}$ = $0.016$ K, $\vartheta_{\pm}$ = $-0.003$ K, $\vartheta_{\pm\pm}$ = $-0.018$ K, and $\vartheta_{z \pm}$ = $0.012$ K (with an approximate error of 3.7\%) --- that successfully reproduced  the ground state spin dynamics (see Fig.~\ref{Fig2:GroundState}\textbf{(h)}). 
The exchange parameters of the $J=15/2$ manifold ($\vartheta_{zz}$, $\vartheta_{\pm}$, etc.) are related to their effective spin-1/2 counterparts ($J_{zz}$, $J_{\pm}$, etc.) through a projection onto the ground-state doublet governed by the effective g-factors. As detailed in the SM~\cite{SM}, applying this transformation confirms the consistency between our two models.
The consistency is further confirmed by comparing energy cuts at the K and $\Gamma$ points [Fig.~\ref{Fig2:GroundState}\textbf{(g)}], where the SU$\left(N\right)$ calculation (solid lines) accurately reproduces both the experimental data (symbols) and the LSW theory (shaded areas).
Thus, the two models are demonstrably equivalent for describing the ground state excitations.

This framework not only reproduces the dispersive CEF excitations observed in our INS data [Fig.~\ref{Fig3:DisCEF}\textbf{(a), \textbf{(c)}, \textbf{(e)}}] with high  fidelity [Fig.~\ref{Fig3:DisCEF}\textbf{(b), \textbf{(d)}, \textbf{(f)}}], but more importantly, it provides physical insight into their origin.
By capturing the coupling between local CEF states and collective spin excitations, the model reveals how the dipolar-ordered ground state drives the emergence of momentum-dependent CEF modes --- establishing a unified understanding of the low- and high-energy dynamics in rare-earth magnets.
As shown in Fig.~\ref{Fig3:DisCEF}\textbf{(g)} and \textbf{(h)}, high-resolution calculations reveal that all low-lying CEF excitations are dispersive. While such exchange-driven dispersion due to energetic proximity is known, as in the pyrochlore \ce{Er2Ti2O7}~\cite{PhysRevB.97.024415}, our work uncovers a deeper, symmetry-based principle governing the nature of this coupling. Remarkably, we find that CEF excitations sharing the same $\Gamma_{4}$ symmetry as the ground state simply replicate its magnon dispersion, while the symmetry-distinct $\Gamma_{5,6}$ DO  doublet exhibits a qualitatively unique dispersive character.
This distinction arises from their contrasting symmetries: 
both the ground state and the second and third excited doublets transform according to the $\Gamma_{4}$ representation, which carries dipolar character. 
Protected by symmetry, magnon dispersions originating from the dipolar ground state can propagate almost unaltered onto the second and third CEF levels, which share the same symmetry as the ground state. 
In sharp contrast, the first excited DO doublet transforms as $\Gamma_{5,6}$ and carries higher-rank multipolar components.
When the dipolar magnons interact with this level, they become renormalized, giving rise to a new, symmetry-allowed multipolar dispersive mode.
These results underscore the key role of symmetry selection in governing the structure and dynamics of collective magnetic excitations.
In addition, subtle differences emerge among the dipolar ground, second, and third CEF states, particularly in scattering intensity, reflecting variations in their $\left| 15/2, m_{j} \right\rangle$ wave function compositions. 

\begin{figure}[t!]
	\includegraphics[angle=0,width=1\linewidth]{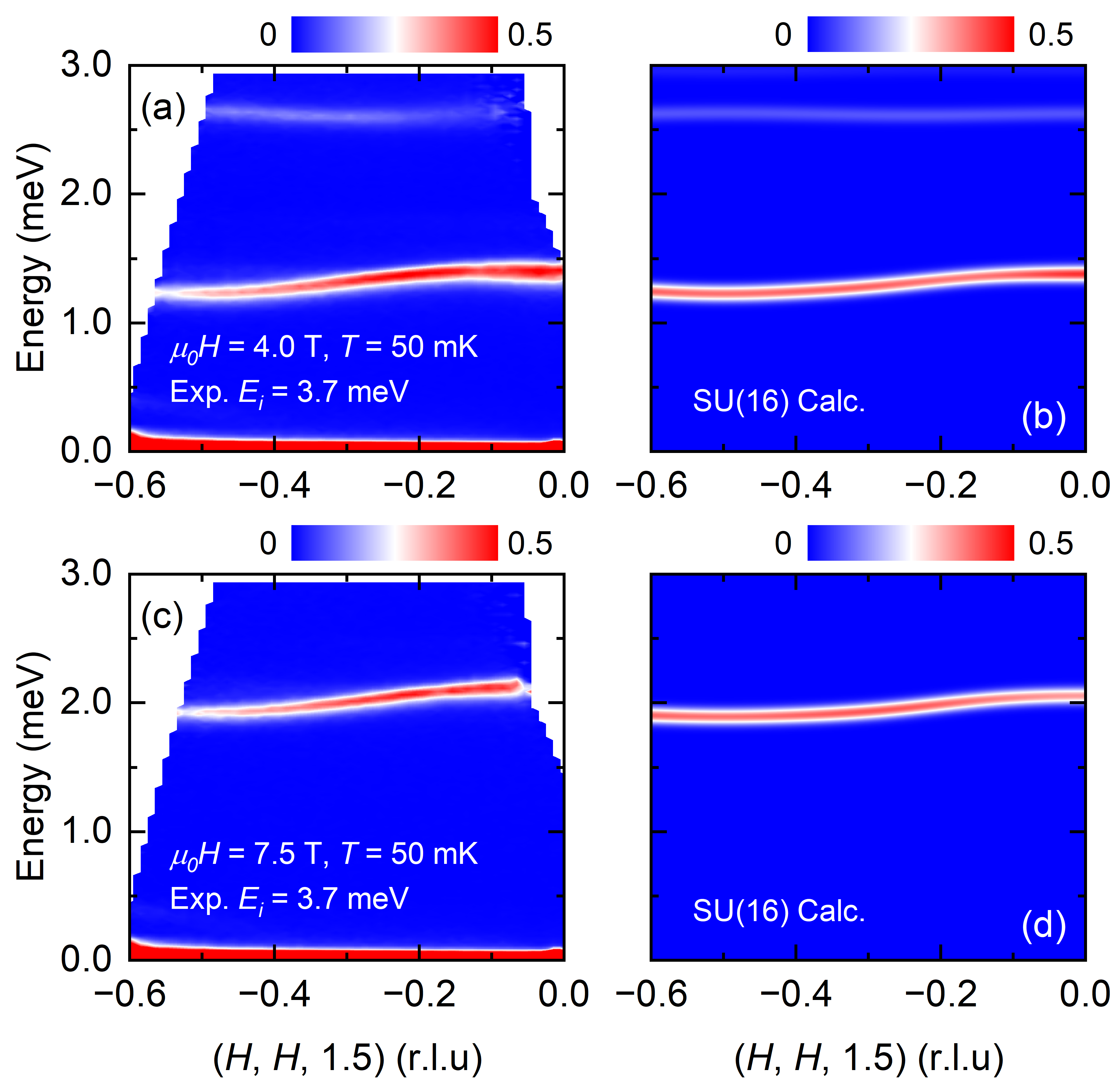}
	\renewcommand{\figurename}{\textbf{Fig. }}
	\caption{\textbf{INS spectra under magnetic fields of 4.0 T and 7.5 T.}
		Panels \textbf{(a, c)} show experimental data along the $\left(H, H, 1.5\right)$ direction at 50 mK, while \textbf{(b, d)} are the corresponding SU$\left(N\right)$ calculations.
	}
	\label{Fig4:FieldEffects}
\end{figure}
\textit{Field-induced multipolar order}---To tune the coupling between local CEF excitations and collective magnetic modes, the magnetic fields are applied to modify the spin dynamics. 
The excitation spectra along the $a$-axis were obtained at 4 T and 7.5 T [Fig.~\ref{Fig4:FieldEffects}\textbf{(a)} and \textbf{(c)}]. 
The field not only lifts the degeneracy of the CEF levels via Zeeman splitting but, more importantly, induces nonlinear shifts that strongly reconstruct the wavefunctions. This reconstruction is directly evidenced by the introduction of significant higher-rank multipolar components, such as $|\pm\frac{15}{2}\rangle$ and $|\pm\frac{9}{2}\rangle$, into the ground state wavefunction which was purely dipolar at zero field~\cite{SM}.

Crucially, the stripe‑$x$ dipolar ground state collapses into a field-polarized ferromagnetic state. 
However, due to the field-induced mixing of higher-rank multipolar components into the wave functions, this ferromagnet carries not only dipolar but also substantial multipolar characters, forming a field-induced multipolar ferromagnet. 
This multipolar polarization is not merely a static property but actively reshapes the magnetic excitation spectrum, modifying the dispersions of both the ground-state modes and the higher CEF excitations.

Field-dependent calculations [Fig.~\ref{Fig4:FieldEffects}\textbf{(b)} and \textbf{(d)}] reproduce these spectral changes with high fidelity, including the field-induced splitting and the emergence of modified dispersive features in the higher CEF levels~\cite{unpublished}. 
These results demonstrate that the magnetic field not only reorients the long-range dipolar order but also stabilizes a multipolar-polarized ground state that reorganizes the collective dynamics.
The stability of this state is confirmed at 7.5 T, where the multipolar character of the wavefunctions is found to be approaching saturation, highlighting the breakdown of the effective spin-1/2 model in this high-field regime~\cite{SM}.
This establishes \ce{NaErSe2} as an ideal platform where both ground-state symmetry and excitation spectrum structure can be continuously tuned by external fields, offering new opportunities for controlling multipolar magnetism in frustrated systems.

\textit{Summary}---In the triangular magnet \ce{NaErSe2}, we identify emergent multipolar excitations arising from the coupling between local CEF modes and collective magnons. Using neutron scattering and a total angular momentum $J=15/2$ model, we demonstrate that a dipolar stripe-$x$ ground state drives the momentum-dependent dispersion of CEF excitations, particularly a prominent DO mode. Furthermore, an applied magnetic field reconstructs the CEF wavefunctions, inducing a multipolar ferromagnetic phase that globally reshapes this coupled excitation spectrum. These findings establish a framework for understanding and controlling multipolar dynamics in frustrated rare-earth magnets.

\textit{Acknowledgements}---We gratefully acknowledge Prof. Andrej Zorko from Jožef Stefan Institute \& University of Ljubljana for helpful discussion.
We also thank Prof. Junfeng Wang from the Wuhan National High Magnetic Field Center, Huazhong University of Science and Technology for his assistance with the pulsed high magnetic field magnetization measurements.
This work was supported by 
the National Key Research and Development Program of China (Grant Nos. 2022YFA1402700 and 2024YFA1408300), 
the National Science Foundation of China (Grant No. 12274186), the Synergetic Extreme Condition User Facility (SECUF, \href{https://cstr.cn/31123.02.SECUF}{https://cstr.cn/31123.02.SECUF} ), and the Fund for High-level Talent of Lanzhou.


\begin{thebibliography}{56}
\bibitem [{unp()}]{unpublished}%
	\BibitemOpen
	\href@noop {} {}\bibinfo {note} {For \ce{NaErSe2}, the higher-lying {CEF}
		levels exhibit pronounced nonlinear splitting under high magnetic fields,
		which will be discussed in detail in a separate publication}\BibitemShut
	{NoStop}%
\end{thebibliography}
%

\end{document}